\documentclass[amsmath,amssymb,aps,prl,twocolumn,superscriptaddress,nofootinbib]{revtex4-1}
\usepackage{graphicx,color,tikz,pgf,csquotes,hyperref,braket,bbm,tensor,upgreek}
\usepackage{enumitem}
\usepackage[caption=false]{subfig}

\usepackage{tikzit}
\tikzstyle{Node}=[fill=black, draw=black, shape=circle, scale=0.3px]
\tikzstyle{coh}=[fill=white, draw=black, shape=circle, scale=0.6px, line width=1px]
\tikzstyle{coh_big}=[fill=white, draw=black, shape=circle, scale=0.6px]
\tikzstyle{coh_black}=[fill=black, draw=black, shape=circle, scale=0.6px, line width=1px]
\tikzstyle{coh_blue}=[fill=blue, draw=blue, shape=circle, tikzit fill=blue, scale=0.5px]
\tikzstyle{coh_wb}=[fill=white, draw=blue, shape=circle, scale=0.6px, line width=1px]

\tikzstyle{line}=[-, fill=none, line width=1px]
\tikzstyle{blockline}=[-, fill=black, line width=3px]
\tikzstyle{boost}=[-, fill={rgb,255: red,128; green,128; blue,128}, draw={rgb,255: red,128; green,128; blue,128}, tikzit fill={rgb,255: red,128; green,128; blue,128}, tikzit draw={rgb,255: red,128; green,128; blue,128}, line width=5px]
\tikzstyle{specialsu2}=[-, line width=5px, fill=black, draw={rgb,255: red,0; green,0; blue,189}, tikzit fill=white, tikzit draw=black]
\tikzstyle{boxdash}=[-, line width=5px, fill=black, dash pattern=on 2pt off 2pt]
\tikzstyle{arrow}=[line width=1.1px, ->]
\tikzstyle{arrowdotted}=[line width=1.1px, ->, dash pattern=on 2pt off 1.5pt]
\tikzstyle{dashed}=[-, line width=1px, dash pattern=on 2pt off 1pt, fill=none]
\tikzstyle{thindash}=[-, line width=0.5px, dash pattern=on 1pt off 3pt, draw={rgb,255: red,158; green,158; blue,158}]
\tikzstyle{grayfill}=[-, fill={rgb,255: red,234; green,234; blue,234}, draw=none]
\tikzstyle{thingray}=[-, line width=0.8px, draw={rgb,255: red,158; green,158; blue,158}]
\tikzstyle{arrowgray}=[draw={rgb,255: red,158; green,158; blue,158}, ->, line width=1px]
\tikzstyle{line_blue}=[-, line width=1.5px, draw=blue, tikzit draw=blue]
\tikzstyle{blue_dashed}=[-, draw=blue, tikzit draw=blue, line width=1.5px, dash pattern=on 2pt off 1.5pt]

\usetikzlibrary{shapes.geometric}

\usepackage{braket}
\renewcommand\bra[1]{{\langle{#1}|}}
\makeatletter
\renewcommand\ket[1]{%
  \@ifnextchar\bra{\k@t{#1}\!}{\k@t{#1}}%
}
\newcommand\k@t[1]{{|{#1}\rangle}}
\makeatother

\newcommand{\one}{\mathbbm 1}
\newcommand{\R}{\mathbb R}

\newcommand{\e}{\textrm{e}}
\newcommand{\SUO}{\mathrm{SU}(1,1)}

\begin{document}

\title{Partial absence of cosine problem in 3d Lorentzian spin foams}

\author{Alexander F. Jercher}
\email{alexander.jercher@uni-jena.de}
\affiliation{Theoretisch-Physikalisches Institut, Friedrich-Schiller-Universit\"{a}t Jena,\\ Max-Wien-Platz~1, 07743 Jena, Germany, EU
}
\affiliation{
Arnold Sommerfeld Center for Theoretical Physics,
Ludwig-Maximilians-Universit\"at M\"unchen,\\
Theresienstr.~37, 
80333 M\"unchen, Germany, EU
}
\affiliation{
Munich Center for Quantum Science and Technology (MCQST),\\ Schellingstr.~4, 
80799 M\"unchen, Germany, EU
}

\author{Jos\'{e} D. Sim\~{a}o}
\email{j.d.simao@uni-jena.de}
\affiliation{Theoretisch-Physikalisches Institut, Friedrich-Schiller-Universit\"{a}t Jena,\\ Max-Wien-Platz~1, 07743 Jena, Germany, EU
}
\author{Sebastian Steinhaus}
\email{sebastian.steinhaus@uni-jena.de}
\affiliation{Theoretisch-Physikalisches Institut, Friedrich-Schiller-Universit\"{a}t Jena,\\ Max-Wien-Platz~1, 07743 Jena, Germany, EU
}

\begin{abstract}
We study the semi-classical limit of the recently proposed coherent spin foam model for (2+1) Lorentzian quantum gravity. Specifically, we analyze the gluing equations derived from the stationary phase approximation of the vertex amplitude. Typically these exhibit two solutions yielding a cosine of the Regge action. However, by inspection of the algebraic equations as well as their geometrical realization, we show in this note that the behavior is more nuanced: when all triangles are either spacelike or timelike, two solutions exist. In any other case, only a single solution is obtained, thus yielding a single Regge exponential.
\end{abstract}

\maketitle

\section{Introduction}

In establishing a connection between spin foam models and gravitational physics, studying the semi-classical limit of the vertex amplitude is a pivotal step. The Engle-Pereira-Rovelli-Livine (EPRL) model~\cite{Engle:2007em,Engle:2008fj,Freidel:2008fv}, the most prominent spin foam model for (3+1)-dimensional Lorentzian quantum gravity, restricts all tetrahedra to be spacelike. In the asymptotic limit, its coherent vertex amplitude~\cite{Livine:2007bq} is dominated by the cosine of the Lorentzian Regge action of a $4$-simplex for appropriately chosen boundary data~\cite{Barrett:2009mw}.

The Conrady-Hnybida (CH) extension~\cite{Conrady:2010kc,Conrady:2010vx} of the EPRL model incorporates spacelike and timelike tetrahedra and triangles. If the interfaces of tetrahedra are spacelike~\cite{Simao:2021qno,Kaminski:2017eew}, the semi-classical approximation of the EPRL-CH vertex amplitude is dominated by a cosine of the Regge action. However, for timelike interfaces, the stationary phase approximation cannot be straightforwardly applied because: (i) the critical points are not isolated~\cite{Liu:2018gfc,Simao:2021qno}\footnote{The action associated to timelike triangles is purely imaginary, such that one lacks the familiar gluing condition for tetrahedra; the spinor variables remain unconstrained. This also occurs in the (2+1) coherent model for spacelike edges~\cite{Simao:2024don}.} (ii) depending on their definition, $\SUO$ coherent states in the continuous series either need to be asymptotically approximated~\cite{Liu:2018gfc} or require a regularization~\cite{Simao:2024don} (iii) the integrand exhibits a branch cut at the critical points~\cite{Simao:2021qno}. A semi-classical formula of the vertex amplitude for boundary states of a $4$-simplex with timelike triangles therefore remains unknown.  

\begin{figure}
    \centering
    \includegraphics[width=0.35\linewidth]{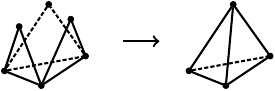}\hspace{1cm}
    \includegraphics[width=0.35\linewidth]{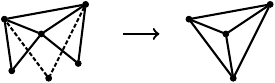}
    \caption{Two ways of folding triangles into a tetrahedron,  leaving one face fixed. The resulting tetrahedra are reflections of one another with respect to the fixed triangle.}
    \label{fig:folding}
\end{figure}

The lack of a semi-classical amplitude for the full set of Lorentzian $4$-dimensional building blocks prompted exploration of the simpler (2+1)-dimensional case. This led to the development of a new coherent state model~\cite{Simao:2024don}, addressing the aforementioned issues. Applying a stationary phase approximation, one obtains two critical point equations corresponding to triangle closure and gluing of triangles along edge vectors. For given boundary data one solution to these equations always exists, corresponding to a geometrical tetrahedron~\cite{Simao:2024don}. A second solution is generally expected to exist, which would correspond to the reflected tetrahedron, obtained by individually rotating triangles. This is graphically represented in Fig.~\ref{fig:folding}.

In this article, we show that a second solution to the critical point equations only exists if all triangles are either spacelike or timelike. This statement is proven from two perspectives. First, we analyze the critical point equations which imply that for certain configurations, the second solution would lie outside the domain of $\SUO$. Second, we argue geometrically that for these configurations, a ``folding'' procedure as depicted in Fig.~\ref{fig:folding} cannot be realized with $\mathrm{SO}(2,1)$-transformations. We conclude from this analysis that for particular causal characters, the semi-classical formula of the (2+1) coherent vertex amplitude is given by the exponential of the Regge action rather than the cosine of it.

\section{Critical point equations}

In this section, we briefly outline the construction and asymptotic analysis of the (2+1) coherent model and refer the reader to~\cite{Simao:2024don} for a detailed treatise. The derivation of quantum amplitudes heavily relies on the representation theory of $\SUO$~\cite{Simao:2024zmv,Bargmann:1946me,rühl1970lorentz} and the utilization of $\SUO$ coherent states~\cite{Perelomov:1986tf,Simao:2024zmv}.

The coherent vertex amplitude is constructed as a convolution of $\SUO$-pairings 
\begin{align}
\!{}^{ n}\scalebox{0.65}{\begin{tikzpicture}
	\begin{pgfonlayer}{nodelayer}
		\node [style=coh] (1) at (-0.75, 0.25) {};
		\node [style=coh] (5) at (0.75, 0.25) {};
		\node [style=none] (6) at (0, 0.5) {};
		\node [style=none] (7) at (0, 0) {};
	\end{pgfonlayer}
	\begin{pgfonlayer}{edgelayer}
		\draw [style=line] (1) to (5);
		\draw [style=blockline] (6.center) to (7.center);
	\end{pgfonlayer}
\end{tikzpicture}
}^{ n'}&:= d_k \braket{qk,n|D^{k(q)}(g)|qk,n'}\,,\label{eq:tl pairing}\\[7pt]
\!{}^{ n}\scalebox{0.65}{\begin{tikzpicture}
	\begin{pgfonlayer}{nodelayer}
		\node [style={coh_black}] (1) at (-0.75, 0.25) {};
		\node [style={coh_black}] (5) at (0.75, 0.25) {};
		\node [style=none] (6) at (0, 0.5) {};
		\node [style=none] (7) at (0, 0) {};
	\end{pgfonlayer}
	\begin{pgfonlayer}{edgelayer}
		\draw [style=line] (1) to (5);
		\draw [style=blockline] (6.center) to (7.center);
	\end{pgfonlayer}
\end{tikzpicture}}^{ n'}&:= d_s \,  \mathcal{C}_{n,gn'} \; [ j,n|D^{j(0)}(g)|j,n'\rangle\,,\label{eq:sl pairing}
\end{align}
where $D^{k(q)}, D^{j (\delta)}$ are $\SUO$-Wigner matrices in the discrete series, $k\in -\frac{\mathbb{N}}{2}$, $q=\pm$, and the continuous series, $j\in-\frac{1}{2}+i\R_+$, $\delta=0$, respectively. Here, $|qk,n\rangle$ is a coherent state in the $L^3$ eigenbasis and $|j,n\rangle$ and $\vert j,n]$ are coherent states in the $K^2$ eigenbasis~\cite{Simao:2024don}. $d_s= s \tanh \pi s $ and $d_k=-2k-1$ are factors arising from the $\SUO$-Plancherel measure. Geometrically, states in the discrete series are associated to timelike edge vectors while states in the continuous series are associated to spacelike edge vectors.

Two novelties that have been introduced in~\cite{Simao:2024don} enter Eq.~\eqref{eq:sl pairing}. First, the pairings of coherent states in the continuous series have been regularized and therefore satisfy the required scaling property, $\!\scalebox{0.65}{\begin{tikzpicture}
	\begin{pgfonlayer}{nodelayer}
		\node [style={coh_black}] (1) at (-0.75, 0.25) {};
		\node [style={coh_black}] (5) at (0.75, 0.25) {};
		\node [style=none] (6) at (0, 0.5) {};
		\node [style=none] (7) at (0, 0) {};
	\end{pgfonlayer}
	\begin{pgfonlayer}{edgelayer}
		\draw [style=line] (1) to (5);
		\draw [style=blockline] (6.center) to (7.center);
	\end{pgfonlayer}
\end{tikzpicture}} \sim (\cdot)^j$, for the asymptotic analysis. Second, $\mathcal{C}_{nn'}$ is a Gaussian that has been added by hand to implement the gluing constraint of spacelike edges, thus guaranteeing a well-behaved semi-classical limit.   

By providing the boundary data of 12 group elements $n_{ab}$ and 6 spin labels $j_{ab} = j_{ba}$, interpreted as edge vectors and their length, the vertex amplitude is pictorially represented by
\begin{equation}
\label{diag}
\mathcal{A}_v(j_{ab},n_{ab}) = \scalebox{0.35}{ \begin{tikzpicture}
	\begin{pgfonlayer}{nodelayer}
		\node [style={coh_black}] (1) at (-0.75, 2.75) {};
		\node [style={coh_black}] (2) at (0, 2.75) {};
		\node [style={coh_black}] (3) at (0.75, 2.75) {};
		\node [style=none] (20) at (-1.25, 2) {};
		\node [style=none] (21) at (1.25, 2) {};
		\node [style=coh] (60) at (-0.75, -2.25) {};
		\node [style={coh_black}] (61) at (0, -2.25) {};
		\node [style=coh] (62) at (0.75, -2.25) {};
		\node [style=coh] (63) at (-2.5, -0.5) {};
		\node [style=coh] (64) at (-2.5, 0.25) {};
		\node [style={coh_black}] (65) at (-2.5, 1) {};
		\node [style=coh] (66) at (2.5, -0.5) {};
		\node [style=coh] (67) at (2.5, 0.25) {};
		\node [style={coh_black}] (68) at (2.5, 1) {};
		\node [style=none] (69) at (-1.25, -1.5) {};
		\node [style=none] (70) at (1.25, -1.5) {};
		\node [style=none] (71) at (1.75, 1.5) {};
		\node [style=none] (72) at (1.75, -1) {};
		\node [style=none] (73) at (-1.75, 1.5) {};
		\node [style=none] (74) at (-1.75, -1) {};
	\end{pgfonlayer}
	\begin{pgfonlayer}{edgelayer}
		\draw [style=blockline] (20.center) to (21.center);
		\draw [style=blockline] (21.center) to (20.center);
		\draw [style=blockline] (69.center) to (70.center);
		\draw [style=blockline] (70.center) to (69.center);
		\draw [style=blockline] (72.center) to (71.center);
		\draw [style=blockline] (74.center) to (73.center);
		\draw [style=line] (2) to (61);
		\draw [style=line] (64) to (67);
		\draw [style=line, bend right=45, looseness=1.75] (3) to (68);
		\draw [style=line, bend left=45, looseness=1.75] (1) to (65);
		\draw [style=line, bend left=45, looseness=1.75] (63) to (60);
		\draw [style=line, bend right=45, looseness=1.75] (66) to (62);
	\end{pgfonlayer}
\end{tikzpicture}},
\end{equation}
where solid lines indicate $\SUO$ integrations. As an example we chose here three spacelike edges (black dots) in the same triangle and three timelike edges (white dots). 

The asymptotic behavior of the vertex amplitude is obtained via a stationary phase approximation~\cite{Barrett:2009ci,Conrady:2008mk,Kaminski:2017eew,Liu:2018gfc,Simao:2021qno,Simao:2024don}. To that end, the spins are uniformly re-scaled, $j_{ab}\rightarrow\Lambda j_{ab}$, and the vertex amplitude is re-written as
\begin{equation}
\mathcal{A}_v = \int\prod_{a=1}^3\mathrm{d}{g_a}\prod_{a<b}f_{ab}( \Lambda j_{ab},n_{ab})\e^{\Lambda S_{ab}}.
\end{equation}
It follows from Hörmander's theorem~\cite[Th. 7.7.5]{Hormander2003} that in the limit $\Lambda \rightarrow \infty$ the integral is dominated by stationary contributions with maximal real part. This yields two sets of equations, referred to as \textit{gluing} and \textit{closure} equations, respectively. 

Using the canonical spin homomorphism $\pi:\SUO\longrightarrow\mathrm{SO}(2,1)$, the geometrical meaning of gluing and closure equations is made explicit. To that end, define spacelike and timelike normal vectors as $v_{ab}^{\mathrm{sl}}:=\pi(n_{ab}^{\mathrm{sl}})\hat{e}_2$ and $v_{ab}^{\mathrm{tl}}=\pi(n_{ab}^{\mathrm{tl}})\hat{e}_0$ where $n_{ab}^{\mathrm{sl}}$ acts on states of continuous series while $n_{ab}^{\mathrm{tl}}$ acts on states of the discrete series. Then, the closure condition reads
\begin{equation}
\label{closuresu11}
  \forall b\,,\; \sum_{a|a\neq b}s_{ab} \epsilon_{ab} v_{ab} =0 \,,
\end{equation}
where $\epsilon_{ab}$ is a sign $\pm$ which depends on the orientations of the diagram \eqref{diag}. Gluing is expressed as
\begin{equation}\label{eq:gluing}
\pi(g_a)v_{ab} = \pi(g_b)v_{ba}.
\end{equation}
Provided that the boundary data corresponds to a geometrical tetrahedron with identified edges, $v_{ab}=v_{ba}$, the gluing equations are solved by $g_a=\one$ for all $a$. Gauge fixing $g_{\bar{a}} = \one$, it has been shown in the Appendix of~\cite{Simao:2024don} that, if existent, a second solution is given by
\begin{equation}\label{eq:g_a sol}
g_b = e^{-i \theta_{\bar{a}b}  \sigma_3\,  (v_{\bar{a}b}\cdot \varsigma)\,\sigma_3 },\quad \theta_{\bar{a}b}\in\R,
\end{equation}
with $\varsigma:=(\sigma_3, i\sigma_2, -i\sigma_1)$ and the dot $(\cdot)$ denoting the scalar product with respect to the Minkowski metric $\eta = \mathrm{diag}(1,-1,-1)$. The dihedral angles $\theta_{ac}$ computed from the normalized edge vectors are given by
\begin{align}
\theta_{ac}&=\mathrm{arctanh}\, \frac{v_{cb} \cdot v_{ab}\times v_{ac}}{(v_{ac}\times v_{cb})\cdot(v_{ab}\times v_{ac})},\quad (ac)\text{ sl},\\[7pt]
\theta_{ac}&=\mathrm{arctan}\, \frac{v_{cb} \cdot v_{ab}\times v_{ac}}{(v_{ac}\times v_{cb})\cdot(v_{ab}\times v_{ac})}\phantom{h},\quad (ac)\text{ tl},
\end{align}
depending on the causal character of the edge $(ac)$. The vector product $(\times)$ still refers to the Minkowski metric. The second solution is obtained by unfolding three of the triangles and gluing them together to form the reflected tetrahedron. A schematic representation is given in Fig.~\ref{fig:folding}.

At this point, we make the main observation of this note: a second solution to the gluing equations does not exist generically. If the boundary data corresponds to a geometrical tetrahedron with identified edge vectors, $v_{ab} = v_{ba}$, then $g_a=\one$ is a solution. However, the second solution requires the rotation of the edge vectors via $\SUO$ transformations given in Eq.~\eqref{eq:g_a sol}. Since these act transitively only on the sheets of the respective hyperboloids, the second solution - otherwise corresponding to an identification of the edges into a reflected tetrahedron - may be obstructed by the presence of light cones. The geometrical reasoning is based on the algebraic equations: if the angle $\theta_{ac}$ has an imaginary part, then $g_b$ in Eq.~\eqref{eq:g_a sol} is not an element of $\SUO$ and the critical point equations are violated. Following the definitions in~\cite{Simao:2021qno,Sorkin:2019llw}, this is precisely the case if the Lorentzian dihedral angle overlaps one or more light cones in the plane $\R^{1,1}$.

In the next section, we show that a second solution exists if either all edges and triangles are spacelike or if all triangles are timelike. Then, the vertex amplitude asymptotes to~\cite{Simao:2024don}
\begin{equation}\label{eq:cos(S)}
\mathcal{A}_v\underset{\Lambda\rightarrow\infty}{\longrightarrow}\Lambda^{\frac{3}{2}}\left(\frac{1}{\sqrt{H_\one}}+\frac{\e^{i 2 S_{\mathrm{R}}}}{\sqrt{H_\theta}}\right),
\end{equation}
where $H_\one$ and $H_\theta$ are the Hessian determinants associated to the first, respectively second critical point solutions. $S_R$ is the Lorentzian Regge action for the given boundary data. Exploiting that coherent states are only defined up to a phase~\cite{Perelomov:1986tf,Simao:2024zmv,Livine:2007bq}, one can consistently multiply this expression by $\e^{-iS_{\mathrm{R}}}$ to find a cosine-like semi-classical formula for the vertex amplitude.\footnote{Because of the additional constraints $\mathcal{C}_{n,n'}$, the Hessian determinants are not simply related via complex conjugation as in~\cite{Dona:2017dvf} and thus, Eq.~\eqref{eq:cos(S)} cannot be expressed as $\cos(S_\mathrm{R})$.}

If the causal character of edges and triangles is different than specified above, only the identity solution exists. Performing the same re-phasing as above, the asymptotic formula of the vertex amplitude is then given by
\begin{equation}
\mathcal{A}_v\underset{\Lambda\rightarrow\infty}{\longrightarrow}\Lambda^{\frac{3}{2}}\frac{\e^{-iS_{\mathrm{R}}}}{\sqrt{H_\one}}.
\end{equation}

We conclude that in (2+1) dimensions, the causal structure of Minkowski space can pose obstructions for obtaining a second solution of the critical point equations. As a result, the asymptotic formula of the vertex amplitude is not given by a cosine of the Regge action but rather a single exponential. This is in contrast to the Euclidean case, where no such obstructions exist and a cosine is obtained generically.

In the next section, we develop a geometrical understanding of the algebraic gluing equations which will lead us to a classification of those case for which a second solution exists.

\section{Geometric realization}

The existence of a second critical point requires the gluing equations~\eqref{eq:gluing} to hold for all $a,b\in\{1,\dots,4\}$. Therefore, we analyze this equation locally, i.e. at a given vertex with an incident triple of faces $(a,b,c)$ and vectors $(v_{ab},v_{ac},v_{bc})$. The boundary data is chosen to correspond to a geometrical tetrahedron with identified vectors $v_{ab} = v_{ba}$. Furthermore, we fix the face $a$ by setting $g_a = \one$.  As a result of gauge-fixing, two of the three equations are given by 
\begin{equation}
    v_{ab} = \pi(g_b) v_{ab},\qquad v_{ac} = \pi(g_c) v_{ac}.
\end{equation}
This can be re-phrased as $\pi(g_b)$ lying in the stabilizer subgroup of $v_{ab}$, which we denote as $S_{v_{ab}}$. Similarly, $\pi(g_c)\in S_{v_{ac}}$. In geometrical terms, a solution to the remaining equation
\begin{equation}\label{eq:3rd gluing}
\pi(g_b) v_{bc} = \pi(g_c) v_{bc}
\end{equation}
exists if the orbits of $v_{bc}$ under the action of $S_{v_{ab}}$ and $S_{v_{ac}}$ have at least one intersection. These orbit spaces are denoted as $O_{v_{bc}}(S_{v_{ab}})$ and $O_{v_{bc}}(S_{v_{ac}})$, respectively. We emphasize that the following arguments apply to any (2+1)-dimensional model which satisfy the gluing equations~\ref{eq:gluing}. This includes in particular models defined on cellular complexes that are more general than triangulations.

In (2+1) dimensions, the plane orthogonal to a spacelike vector $w$ is (1+1)-dimensional Minkowski space, $\R^{1,1}$. Thus, the stabilizer subgroup of  $w$ is isomorphic to $\mathrm{SO}(1,1)$. For $w$ timelike, the orthogonal plane is Euclidean $\R^2$ and therefore, the corresponding stabilizer subgroup is isomorphic to $\mathrm{SO}(2)$. Acting with the stabilizer subgroup of $w$ on another vector $v$ traces out a one-dimensional curve that lies in the plane orthogonal to $w$. The shape of the orbit space $O_v(S_w)$ is thus determined by the causal character of the vectors $v,w$ and the face spanned by the two. More precisely, one finds that
\begin{itemize}
    \item if $w$ is timelike then $O_v(S_w)\cong S^1$.
    \item if $v,w$ are spacelike and the spanned face is spacelike then $v-w$ is a spacelike vector. As a result, the action of $S_w$ on $v$ traces out a spacelike hyperbola, $O_v(S_w)\cong\mathrm{H}_{\mathrm{sl}}^\pm$. Importantly, since the orbit is a subspace of $\R^{1,1}$, the spacelike hyperbola is either of the two disconnected components, indicated with ``$\pm$''. 
    \item if $v,w$ are spacelike and the face they are spanning is timelike, then $v-w$ is a timelike vector. Consequently, the action of $S_w$ on $v$ traces out a timelike hyperbola, $O_v(S_w)\cong \mathrm{H}_{\mathrm{tl}}^\pm$. Also, if $w$ is spacelike and $v$ is timelike the orbit is a timelike hyperbola in $\R^{1,1}$.
\end{itemize}
Two solutions of Eq.~\eqref{eq:3rd gluing} exist if the orbit spaces $O_{v_{bc}}(S_{v_{ab}})$ and $O_{v_{bc}}(S_{v_{ac}})$ intersect twice, that is, if the curves traces out by acting with $S_{v_{ab}}$ and $S_{v_{ac}}$ on $v_{bc}$ intersect twice.

By geometrically constructing these orbit spaces and exploring the space of configurations, we obtain a classification for the number of solutions to the three gluing equations.\footnote{A Mathematica notebook for this construction can be found at \href{https://github.com/Jercheal/No_cosine}{github.com/Jercheal/No cosine}, where an exhaustive collection of \textsc{Manipulate}-plots is provided that allows for the exploration of the configuration space.} It can be entirely captured by the different causal character of the vectors and faces for which an exhaustive list is given in Table~\ref{tab:sols}.  

\begin{table}[]
    \centering
    \begin{tabular}{| c | c | c | c |}
    \hline
    $\quad a\quad$   &   $\quad b\quad $   &   $\quad c\quad$   &   \# sol.\\
    \hline
    \multicolumn{4}{| c |}{$v_{ab},v_{ac},v_{bc}$ spacelike}\\
    \hline
    sl  &   sl  &   sl  &   2  \\
    sl  &   sl  &   tl  &   1  \\
    tl  &   sl  &   sl  &   1  \\
    sl  &   tl  &   tl  &   1   \\
    tl  &   sl  &   tl  &   1   \\
    tl  &   tl  &   tl  &   2   \\
    \hline
    \multicolumn{4}{| c |}{$v_{ab},v_{ac}$ spacelike, $v_{bc}$ timelike}\\
    \hline
    sl  &   tl  &   tl  &   1   \\
    tl  &   tl  &   tl  &   2   \\
    \hline
    \multicolumn{4}{| c |}{$v_{ab},v_{bc}$ spacelike, $v_{ac}$ timelike}\\
    \hline
    tl  &   sl  &   tl  &   1   \\
    tl  &   tl  &   tl  &   2   \\
    \hline
    \multicolumn{4}{| c |}{$v_{ab}$ spacelike, $v_{ac},v_{bc}$ timelike}\\
    \hline
    tl  &   tl  &   tl  &   2   \\
    \hline
    \multicolumn{4}{| c |}{$v_{ab},v_{ac},v_{bc}$ timelike}\\
    \hline
    tl  &   tl  &   tl  &   2   \\
    \hline
    \end{tabular}
    \caption{Characterization of the number of solutions to the critical point equations, where the entries ``sl'' and ``tl'' denote whether the corresponding face is spacelike or timelike, respectively.}
    \label{tab:sols}
\end{table}

We finally conclude that two solutions to the critical point equations exist in two cases. First, if all edges and all triangles are spacelike, the two orbit spaces correspond to spacelike hyperbola in different planes that intersect exactly twice. A visualization of this case can be found in Fig.~\ref{fig1}. Second, if all triangles are timelike a second solution exists irrespective of the causal character of the edges. Visualizations of these cases can be found in Figs.~\ref{fig6}-\ref{fig12}. In all the other cases, there is only one intersection of orbits which can be seen from Fig.~\ref{fig:1sol}.

\section{Conclusions}

\begin{figure*}
\subfloat[\label{fig1}]{
\includegraphics[width=0.33\linewidth]{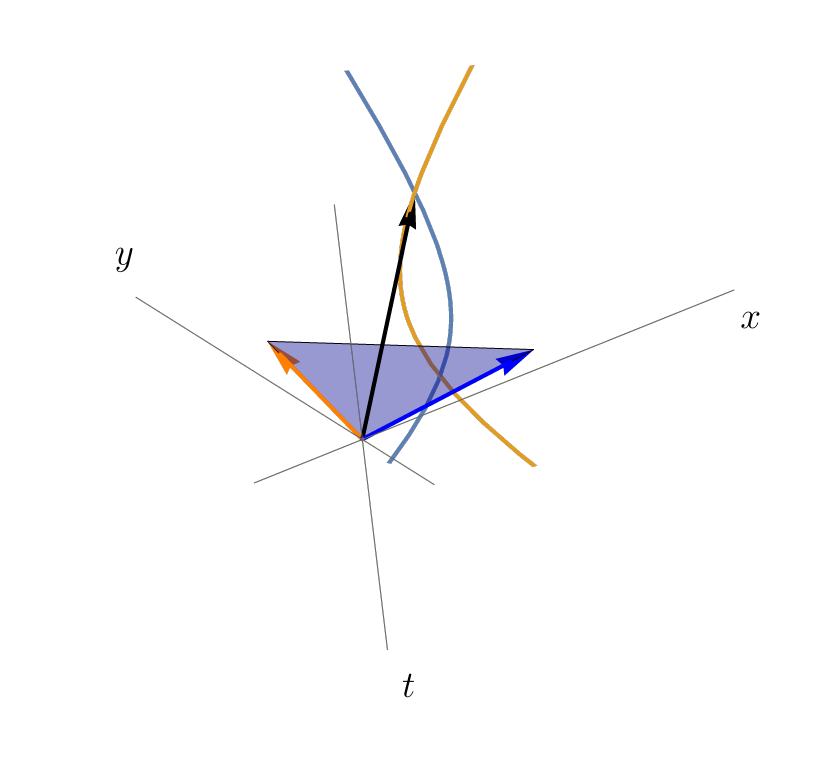}%
}
\subfloat[\label{fig6}]{
\includegraphics[width=0.33\linewidth]{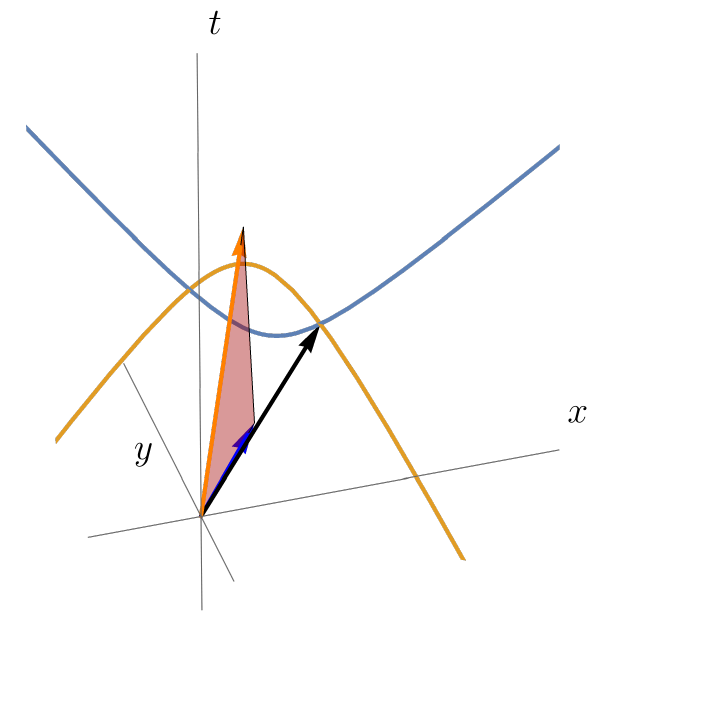}%
}
\subfloat[\label{fig8}]{
\includegraphics[width=0.33\linewidth]{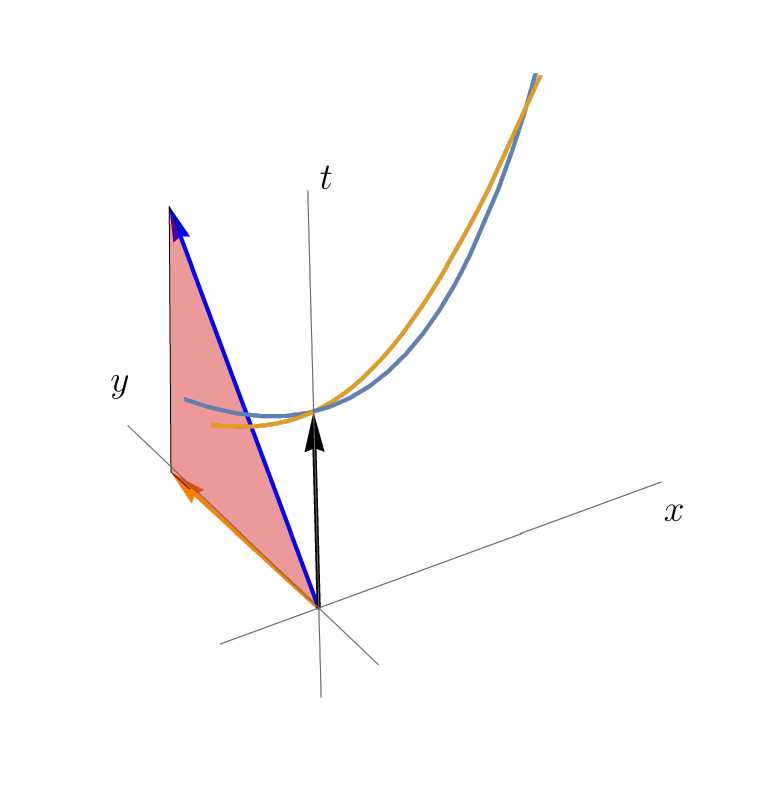}%
}\\
\subfloat[\label{fig10}]{
\includegraphics[width=0.33\linewidth]{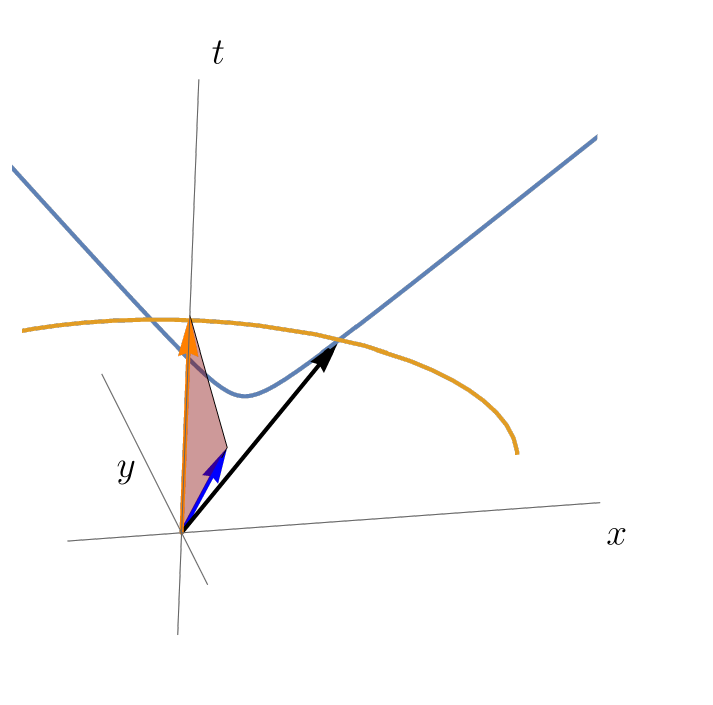}%
}
\subfloat[\label{fig11}]{
\includegraphics[width=0.33\linewidth]{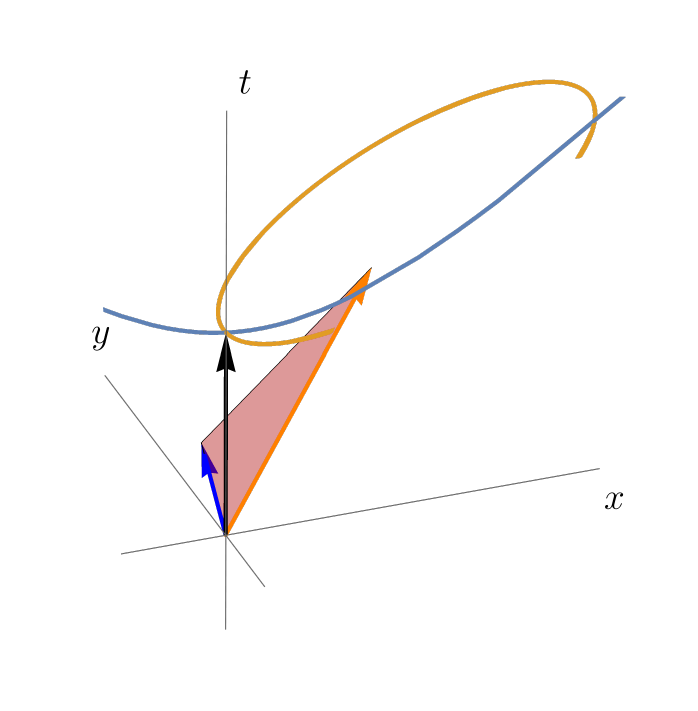}%
}
\subfloat[\label{fig12}]{
\includegraphics[width=0.33\linewidth]{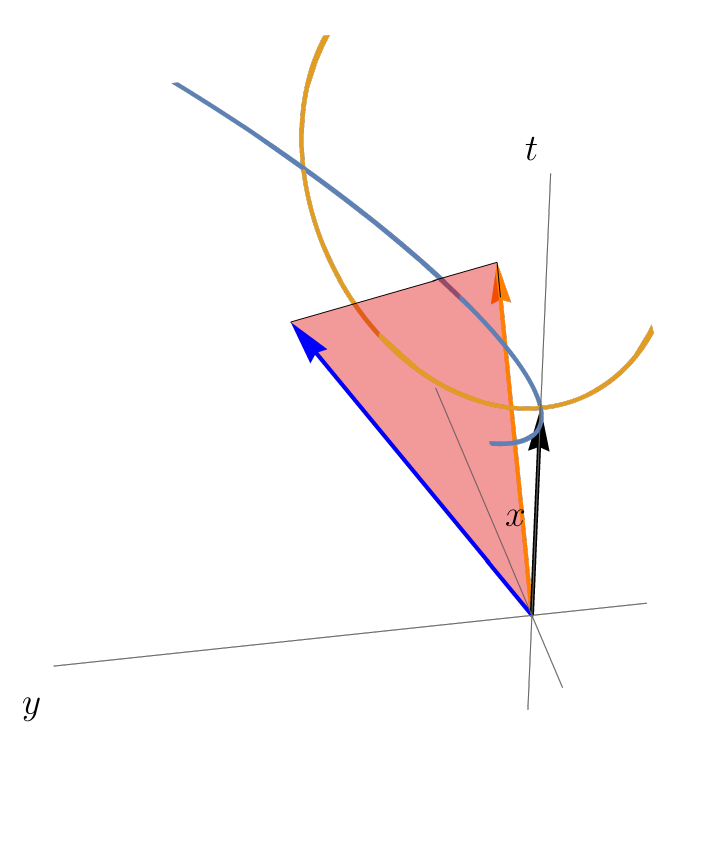}%
}
\caption{The six types of causal configurations that exhibit two solutions. $v_{ab}$ and $v_{ac}$ are drawn as blue and orange vectors while $v_{bc}$ is black. Face $a$ is shared between the blue and orange vector, $b$ is shared between the blue and black vector and $c$ is shared between the orange and black vector. Acting with the stabilizer subgroup of $v_{ab}$ ($v_{ac}$) on the black vector traces out the corresponding orbit, depicted as the blue (orange) curve. The second intersection of the two curves corresponds to the second solution of the gluing equations, obtained by folding faces $b$ and $c$ to the opposite side of the triangle $a$. If the triangle spanned by $v_{ab}$ and $v_{ac}$ is spacelike (timelike) it is drawn as blue (red). (a) All edges and triangles are spacelike. The orbit spaces are two spacelike hyperbola. (b) All edges are spacelike and all faces are timelike. The orbit spaces are two timelike hyperbola. (c) $v_{ab}$ and $v_{ac}$ are spacelike, $v_{bc}$ is timelike and all the faces are timelike. The orbit spaces are two timelike hyperbola. (d) $v_{ab}$ is spacelike, $v_{ac}$ is timelike, $v_{bc}$ is spacelike and all the faces are timelike. The orbit spaces are a timelike hyperbola and a circle. (e) $v_{ab}$ is spacelike, $v_{ac}$ and $v_{bc}$ are timelike and all faces are timelike. The orbit spaces are a timelike hyperbola and a circle. (f) All vectors and all faces are timelike. The orbits are two circles.}
\label{fig:2sols}
\end{figure*}

\begin{figure*}
\subfloat[\label{fig2}]{
\includegraphics[width=0.33\linewidth]{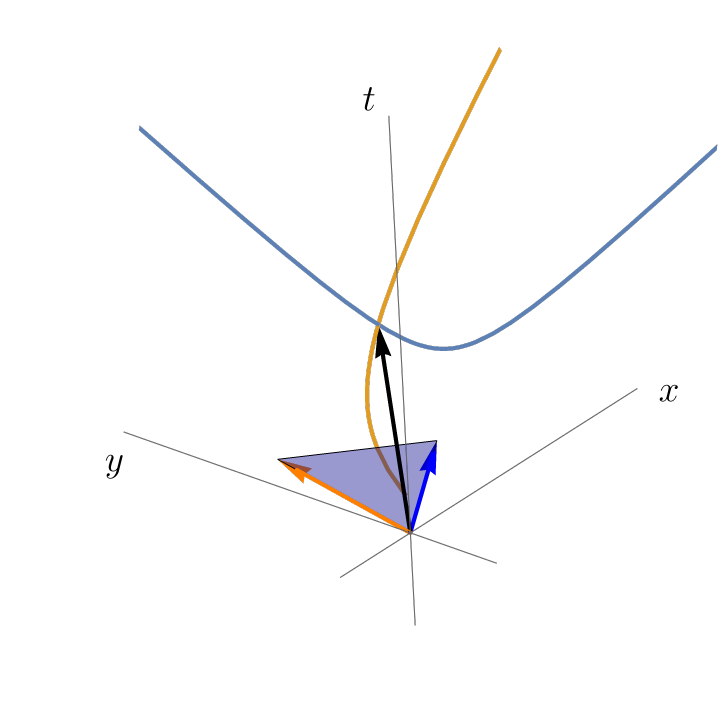}%
}
\subfloat[\label{fig3}]{
\includegraphics[width=0.33\linewidth]{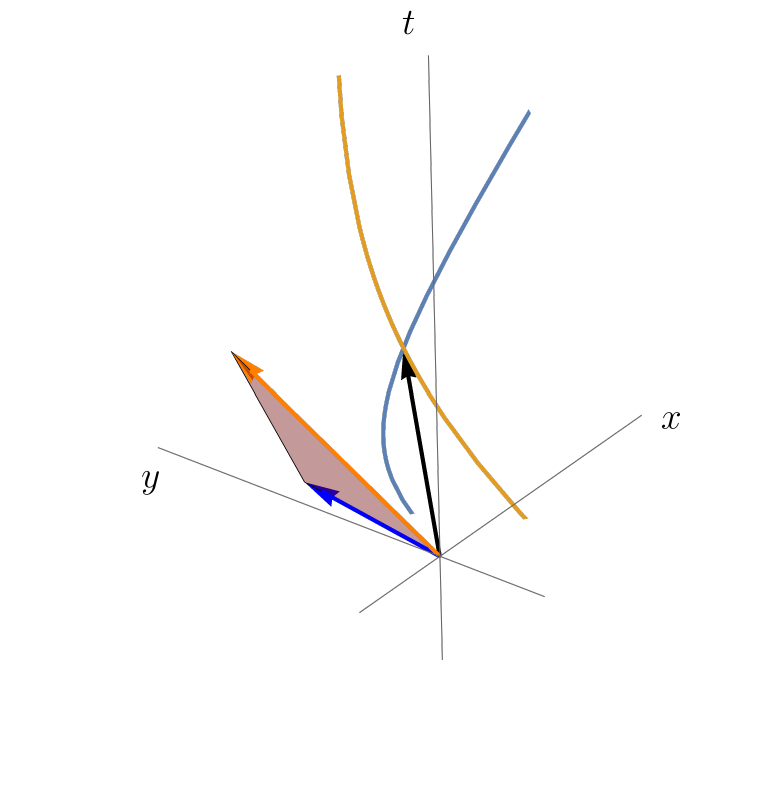}%
}
\subfloat[\label{fig4}]{
\includegraphics[width=0.33\linewidth]{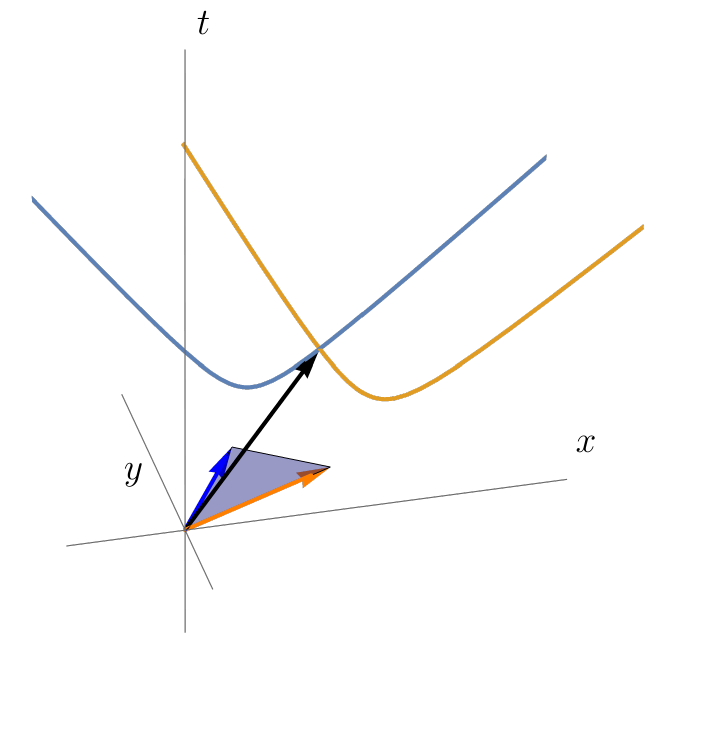}%
}\\
\subfloat[\label{fig5}]{
\includegraphics[width=0.33\linewidth]{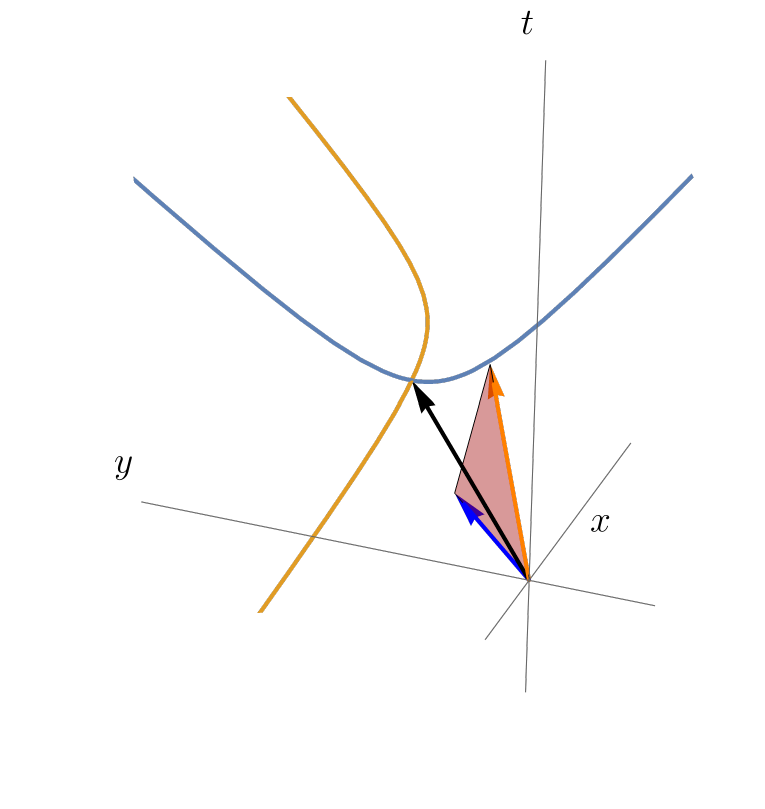}%
}
\subfloat[\label{fig7}]{
\includegraphics[width=0.33\linewidth]{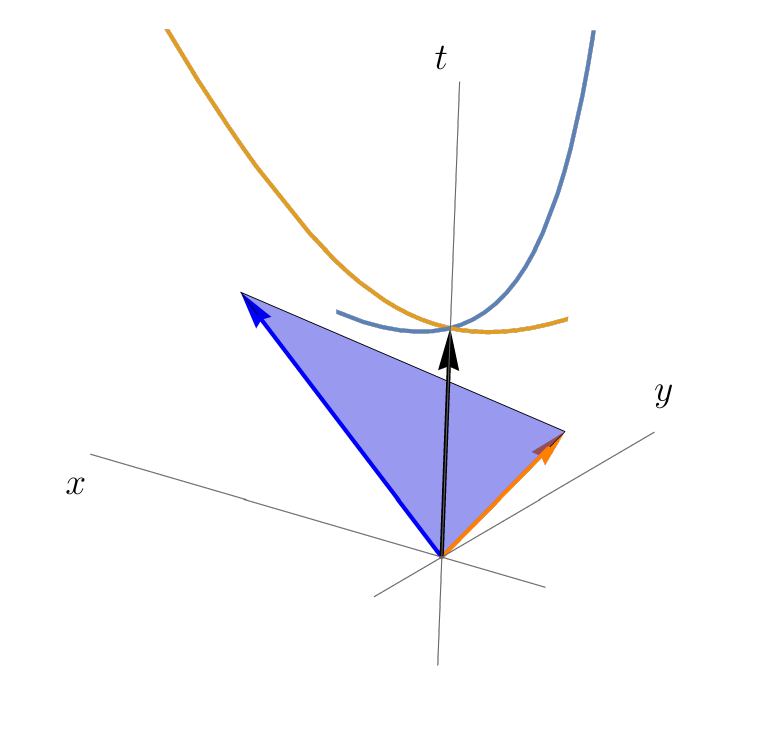}%
}
\subfloat[\label{fig9}]{
\includegraphics[width=0.33\linewidth]{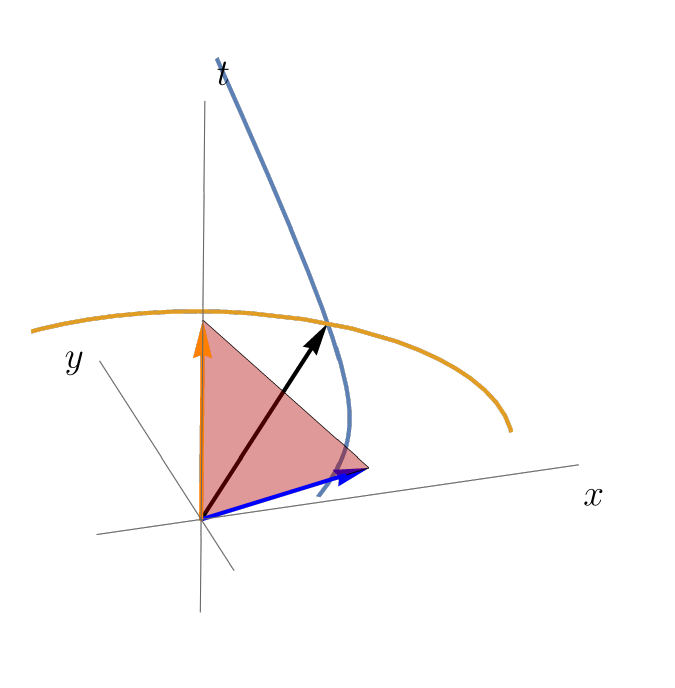}
}
\caption{The six types of causal configurations that exhibit one solution. Vectors, orbits and the triangle are drawn as in Fig.~\ref{fig:2sols}. (a) All vectors are spacelike, the face $b$ is timelike, $a$ and $c$ are spacelike. The orbit spaces are a spacelike and a timelike hyperbola. (b) All vectors are spacelike, the face $a$ is timelike, $b$ and $c$ are spacelike. The orbits are two spacelike hyperbola. (c) All vectors are spacelike, the face $a$ is spacelike, $b$ and $c$ are timelike. The orbits are two timelike hyperbola. (d) All vectors are spacelike. Faces $a$ and $b$ are timelike, $c$ is spacelike. The orbits are a spacelike and a timelike hyperbola. (e) $v_{ab}$ and $v_{ac}$ are spacelike, $v_{bc}$ is timelike, $a$ is spacelike and $b,c$ are timelike. The orbits are two timelike hyperbola. (f) $v_{ab}$ is spacelike, $v_{ac}$ is timelike, $v_{bc}$ is spacelike, the faces $a,c$ are timelike and $b$ is spacelike. The orbits are given by a spacelike hyperbola and a circle.}
\label{fig:1sol}
\end{figure*}

We have shown in this note that the asymptotic limit of (2+1)-dimensional coherent spin foams does not generically lead to a cosine of the Regge action. A cosine is obtained if either all triangles and edges are spacelike or if all triangles are timelike. In all other cases, the vertex amplitude is approximated by a single oscillating Regge exponential. 
\\
\\
\indent In~\cite{Livine:2002rh,Bianchi:2021ric}, the presence of a cosine in the semi-classical approximation of spin foams has been interpreted as an averaging over time orientation. Therein, the proposals for time oriented models are characterized by an a posteriori restriction of the quantum amplitudes, such that in a semi-classical limit only one critical point is obtained. Although not motivated from time orientation, a similar restriction of the quantum amplitudes has been proposed in the context of proper vertex spin foams~\cite{Engle:2015mra,Engle:2015zqa} to resolve the cosine problem. Our results for the (2+1)-dimensional model show that such a restriction is not necessarily required to obtain a single Regge exponential in the asymptotic limit. Whether this implies that the present model already captures a notion of time orientation is unclear. In particular, further investigation of this question would require a firm understanding of the notion of time orientation in spin foams, which has not been developed so far. 
\\
\\
\indent Our results bear consequences for the investigation of quantum cosmology within the present model. Utilizing the semi-classical amplitude as an effective vertex amplitude, its precise form now depends on the causal character of edges and faces, according to the classification we provided here. In the symmetry reduced setting of cosmology, one makes the following interesting observation: the causal configurations that lead to a cosine of the Regge action are those that constitute causality violations as defined in~\cite{Asante:2021phx}. Causally regular configurations on the other hand are exactly those that yield a single Regge exponential in the semi-classical limit. We leave the development of a physical understanding of this observation to future work. 
\\
\\
\indent We close by discussing to which extent the present results can be transferred to (3+1) coherent models, in particular the EPRL-CH model. A crucial difference to the (2+1) coherent model is that the boundary data is associated to 3-dimensional normal vectors of triangles. Still, there exist two cases which allow for an immediate transfer of the geometrical picture developed here. If every tetrahedron, and thus all triangles and edges, are spacelike, boundary data corresponding to a geometrical $4$-simplex can be provided and the gluing equations describe Lorentz transformations that stabilize 2-dimensional planes spanned by bivectors. Acting with the stabilizer subgroups on the triangle normal vectors which live in $S^2$ therefore traces out two circles, similar to Fig.~\ref{fig12}. These intersect twice, thus yielding a second solution of the critical point equations. Also in the case of all tetrahedra and triangles being timelike, the geometrical picture applies. In this case the action of the stabilizer subgroups on a spacelike triangle normal vector traces out two spacelike hyperbola that intersect twice, similar to Fig.~\ref{fig1}. 

For a mixture of spacelike and timelike tetrahedra that meet at \textit{heterochronal}\footnote{A heterochronal interface lies between a spacelike and a timelike polyhedron.} interfaces, the geometrical picture of the (2+1)-dimensional gluing equations cannot be applied. That is because for the (3+1)-dimensional gluing equations of heterochronal interfaces~\cite{Simao:2021qno},
\begin{equation*}
\pi(g_a)^{\wedge 2}\left((1,\vec{0})\wedge (0,\vec{n}_{ab})\right) = \pi(g_b)^{\wedge 2}\left((\vec{0},1)\wedge (\vec{n}_{ba},0)\right),
\end{equation*}
the boundary normal vectors $\vec{n}_{ab}$ and $\vec{n}_{ba}$ cannot be chosen as identified, which is a direct consequence of the very particular embedding of 3-dimensional triangle normals into 4 dimensions. As a result, the gluing equations do not describe rotations that stabilize 2-dimensional planes. It is the case of heterochronal interfaces for which one might suspect only one solution to the critical point equations by a naive transfer of the results given here. However, the asymptotic analysis of~\cite{Kaminski:2017eew} states and proves that indeed two solutions exist. It is therefore conceivable that the analysis conducted here is simply not applicable to the (3+1)-dimensional case because of the intricacies of the bivector gluing equations, the different type of boundary data and the particular embedding of spacelike and timelike tetrahedra. 

\begin{acknowledgements}
\textbf{Acknowledgements}  The authors thank Pietro Don\`{a} for helpful discussions. AFJ, JDS and SSt gratefully acknowledge support by the Deutsche Forschungsgemeinschaft (DFG, German Research Foundation) Grant No 422809950. AFJ and JDS acknowledge support by the DFG under Grant No 406116891 within the Research Training Group RTG 2522/1. AFJ is grateful for the generous financial support by the MCQST via the seed funding Aost 862981-8 granted to Jibril Ben Achour by the DFG under Germany’s Excellence Strategy – EXC-2111 – 390814868.
\end{acknowledgements}

\bibliography{references.bib} 

\end{document}